%% file: main.tex
\documentclass[a4paper,11pt]{article}

\usepackage{pos}
\usepackage{graphicx}
\usepackage{booktabs}
\usepackage{float}
\usepackage{caption}
\usepackage{amsbsy}

\newcommand{\id}{1\!\!1}
\newcommand{\noi}{\vspace*{0.2cm}\noindent}

\title{Meson thermal masses at different temperatures}
\author*[a]{Sergio Chaves García-Mascaraque}
\author[a,b]{Gert Aarts}
\author[a]{Chris Allton}
\author[a]{Timothy Burns}
\author[c]{Simon Hands}
\author[d]{Benjamin J\"{a}ger}

\affiliation[a]{Department of Physics, Swansea University, Swansea SA2 8PP, United Kingdom}
\affiliation[b]{European Centre for Theoretical Studies in Nuclear Physics and Related Areas (ECT*) \& Fondazione Bruno Kessler Strada delle Tabarelle 286, 38123 Villazzano (TN), Italy}
\affiliation[c]{Department of Mathematical Sciences, University of Liverpool, Liverpool L69 3BX, United Kingdom}
\affiliation[d]{CP3-Origins \& Danish IAS, Department of Mathematics and Computer Science, University of Southern Denmark, 5230, Odense M, Denmark}

\emailAdd{989336@swansea.ac.uk}

\abstract{
    We determine the ground state meson masses at low temperature using simulations with $N_f = 2+1$ flavours 
    of improved Wilson-clover fermions. Subsequently we study the effect of increasing the temperature of the hadron gas, including 
    the transition to the quark-gluon plasma, as well as the restoration of SU(2)$_A$ chiral symmetry. We use the 
    FASTSUM anisotropic,  fixed-scale Generation 2L ensembles and consider mesons with light, strange and charm content.
}

\FullConference{%
  The 38th International Symposium on Lattice Field Theory, LATTICE2021
  26th-30th July, 2021
  Zoom/Gather@Massachusetts Institute of Technology
}

\begin{document}

    \maketitle

    % -- Include some overall introduction of the project
    \input{./sections/intro.tex}

    % -- Include the setup information
    \input{./sections/setup.tex}

    % -- Include the regression section
    \input{./sections/regression.tex}

    % -- Include the results section
    \input{./sections/results.tex}

    % -- Include the conclusion section
    \input{./sections/conclusions.tex}

    % -- Include the bibliography

\end{document}

%% file: sections/intro.tex
\section{Introduction}

The fate of hadrons as the temperature of the hadron gas is increased and the transition to the 
quark-gluon plasma is made, has been a topic of longstanding interest. In principle, nonperturbative 
simulations of lattice QCD should provide theoretical insight, but the analysis is complicated due to 
the difficulty in defining the notions of {\em ground state} and {\em mass} at nonzero temperature. 
This can be resolved by considering hadronic spectral functions, which include thermal mass 
shifts and widths, but their construction is hindered by the finite number of points in the Euclidean 
time direction available for the required analytic continuation. Nevertheless, interesting results 
for light and strange baryons have been obtained \cite{Aarts:2018glk, Aarts:2017rrl}, shedding light on parity
doubling and chiral symmetry restoration.

In this contribution, we present the temperature dependence of masses of mesons. We take the following\
conservative approach: we first determine the masses of the ground states in various channels at low
temperature, using a regression analysis based on the one presented in Ref.\ \cite{Bazavov:2019www}. We 
aim to minimise possible bias by performing variational fits and avoiding cherry-picking. Subsequently 
we extend this analysis to nonzero temperature and attempt to systematically generate comparable estimates 
of ground state masses for all temperatures. While we expect this approach to break down at higher 
temperatures and definitely in the quark-gluon plasma phase, where light mesons no longer exist, we 
demonstrate that it nevertheless provides insight into thermal effects. In particular, we study the 
restoration of the SU(2)$_A$ symmetry as the temperature increases, by comparing the masses of the 
$\rho(770)$ and $a_1(1260)$. Finally, we identify limits to this approach by contrasting results obtained 
with local and smeared sources at high temperature.

%% file: sections/setup.tex
\section{Lattice setup and mesonic correlation functions}

We use the anisotropic FASTSUM ensembles described in detail in Ref.\ \cite{Aarts:2020vyb}, with 
$N_f=2+1$ flavours of clover-improved Wilson fermions. The strange quark mass is at its physical value
\cite{Edwards:2008ja}, but the light quarks are heavier than in nature. Relevant parameters are given in Table
\ref{tab:1}. In the fixed-scale approach, the temperature is varied by changing $N_\tau$, using the 
relation $T=1/(N_\tau a_\tau)$.The lowest (highest) temperature we study here is $T=47$ ($300$)~MeV,
obtained using $N_\tau = 128$ ($20$). 

\begin{table}[h]
    \begin{center}
    \begin{tabular}{ccccccc}
    \hline
    $1/a_\tau$ [GeV] & $a_s$ [fm] & $\xi = a_s / a_\tau$  & $N_s$  & $M_\pi$ [MeV] & $M_\pi L$ & $T_c$ [MeV] \\
    \hline
    $5.997(34)$ & $0.01136(6)$ & $3.453(6)$ & $32$  & $236(2)$ & $4.36$ & $164(2)$ \\
    \hline
    \end{tabular}
    \caption{
        Parameters relevant for the ensembles: $a_\tau$ ($a_s$) is the temporal (spatial) lattice spacing; $\xi$ is the renormalised anisotropy; $N_s$ is the number of points in the spatial direction; $M_\pi$ is the mass of the pion; $T_c$ is the critical temperature, computed using the inflection point of the renormalised chiral condensate \cite{Aarts:2020vyb}.    
    }
    \label{tab:1}
\end{center}
\end{table}

Our analysis focuses on mesonic correlation functions. The mesonic operators are $O(a)$ improved through 
the Symanzik improvement scheme. In the correlation functions the source and sink operators are 
identical. As we are  interested in thermal observables, correlation functions are expressed in a
\textit{time-momentum} representation, with the external momentum set to zero. Consequently, our 
correlators only depend on Euclidean lattice time $\tau$, $C(\tau)$ with $\tau/a_\tau  \in [0, N_\tau)$.
In the simulations, three quark fields are present: $u$, $d$ and $s$, with $u$ and $d$ being degenerate. 
We also have access to observables containing charm ($c$), which is however not included in the fermion
determinant. As a result, we can simulate six flavour combinations: $uu$, $us$, $uc$, $ss$, $sc$ and $cc$. 
We do not calculate disconnected contributions, therefore, we only have access to non-singlet flavour
observables.

\begin{table}[t]
    \begin{center}
    \begin{tabular}{l | cccc }
        \hline
        Channel     & pseudoscalar & vector       & axial-vector         & scalar   \\
        Operator    & $\gamma_5$   & $\gamma_\mu$ & $\gamma_\mu\gamma_5$ & $\id$      \\
        $J^{PC}$    & $0^{-+}$     & $1^{--}$     & $1^{++}$             & $0^{++}$ \\
        \hline
    \end{tabular}
    \caption{
        Channels available in our simulations, indicated by the Dirac matrix and the $J^{PC}$ quantum number.
    }
    \label{ta:channels}
    \end{center}
\end{table}

We have analysed a number of channels: pseudoscalar, vector, axial vector and scalar, see
Table~\ref{ta:channels}. Here we report results for the first three channels only. Besides the 
specification of the flavour combination and operator used, we can choose the type of source used in 
the inversion of the Dirac operator. We have used \textit{local} and \textit{smeared} (using Gaussian
smearing with parameters $N = 100, \kappa= 5.5$) sources. The latter are designed to have  better
overlap with the ground state at zero temperature, but in principle both types should lead to the 
same ground state mass. We construct correlation functions in which either local sources were used 
for both quark propagators, and denote these with $l$-$l$, or smeared sources, denoted with $s$-$s$. 
As a result, we have two different estimates for the same mesonic correlator, for each channel and
flavour combination. 

%% file: sections/regression.tex
\section{Regression analysis} 

\subsection{Spectral decomposition}

Every parametric regression analysis needs a model. Here we start from the simplest Ansatz: a sum of
isolated states, characterised by a mass $M_s$, amplitude $A_s$ and vanishing width (i.e.\ simple poles 
in the mesonic correlator, or delta-functions in the corresponding spectral function). The correlator 
then takes the form 
\begin{equation} \label{eq:specdec}
    C(\tau) = \sum_{s = 0}^{\infty} A_s \cosh\left[ M_s (\tau - 1/(2T) \right].
\end{equation}
In principle, we can fit our estimates of the correlation function to the equation above to
extract the masses $M_s$ and amplitudes $A_s$ of the different states. We refer to the state
with the lowest mass as the \textit{ground state}; extracting its mass is our main objective.
Provided $M_0 \ll  M_{s\neq 0}$, the ground state dominates at large times $\tau > 0$, which at 
nonzero temperature is $\tau/a_\tau\simeq N_\tau/2$. It is noted that Eq.\ (\ref{eq:specdec}) is only
expected to be valid in the low-temperature limit, $T \to 0$. Once the temperature is increased, the 
validity of the model starts being questionable, due to in-medium effects.

To perform a regression analysis, we assume that the correlation functions are, at all temperatures,
described by the model
\begin{equation} \label{eq:model}
    C(\tau) = f(\tau; \theta, N_{\rm tr}) + u,
\end{equation}
where the Ansatz, $f$, is a truncation of Eq.~(\ref{eq:specdec}) at the order $N_{\rm tr}$. It depends
on the parameters $\theta=\{A_s, M_s | s=0,\ldots, N_{\rm tr}\}$, which need to be found; the parameters
$A_s$ and $M_s$ correspond to the ones in Eq.~(\ref{eq:specdec}). The error term, $u$, is assumed normally 
distributed and conditionally independent of $\tau$.

To set up notation, from now on $\hat{C}(\tau)$ is our estimate of the correlation function, which is
estimated by averaging over all configurations at a fixed temperature. Note that the regression analysis
presented in this section is based on the procedure presented in Ref.\ \cite{Bazavov:2019www}.

\subsection{Sources of problems in a regression analysis}

Performing a regression analysis is a nontrivial task. Here we briefly review some of the problems present
in the analysis and discuss solutions to mitigate them.

\noi
{\bf Correlation function data is heavily correlated} --
The correlation function estimate at different times $\tau$ is heavily correlated, hence the data includes
less information than expected. The source of this autocorrelation is the computation of the propagators; 
the entire ensemble is used to produce an estimate of the correlation function at all $\tau$. Consequently,
the samples are correlated at different $\tau$, which violates the usual assumption of independence in the
errors in Eq.\ (\ref{eq:model}).

Not taking into account the correlation in the regression analysis tends to underestimate the
uncertainty in the parameter estimates and even produce wrong results. One way to take correlation
into account is by using the so-called \textit{correlated maximum likelihood estimate}. This
estimate can be derived by maximising the likelihood function of the intersection of $N$ correlated 
normally distributed random variables. The problem is equivalent to minimising the following objective 
function:

\begin{equation} \label{eq:chisq}
    \chi^2 = \left[ \hat{C} - f(\tau; \theta)\right]  K^{-1}
             \left[ \hat{C} - f(\tau; \theta)\right]^T,
\end{equation}
$K$ being the covariance matrix of the error term, which can be estimated using the correlation
function estimate $\hat{C}(\tau)$. Minimising Eq.~(\ref{eq:chisq}) tends to be unstable; starting at
different initial parameters leads to significantly dissimilar results.

One can always try to break down the correlation by producing a bootstrap estimate of the data. This
can be done by selecting different random samples from all configurations available for each time
$\tau$. This procedure can produce biased estimates of $\hat{C}(\tau)$ \cite{Efron:1986hys}.

\noi{\bf Multistate fits are a must} --
As explained before, the lowest state tends to dominate the correlation function at large $\tau$. However, 
without having information about the values of the masses of the excited states, we do not know when those 
states start becoming relevant as $\tau \searrow 0$. As a consequence, if only one-state fits are 
produced, then bias is included in the result. Hence one should always try including as many states as
possible in the model.

\noi{\bf Varying the fit windows} --
The fit window used in the regression should be varied to produce as many estimates as possible of the
ground state mass. Varying the fit window reduces the bias included by manually selecting a region of fit.
For the remainder of the contribution, we define a fit window $FW(\tau_0, \tau_f)$ as all times $\tau$ 
included in the interval $[\tau_0, \tau_f]$. We fix $\tau_f/a_\tau = N_\tau/2$.

\noi{\bf Parameter initialisation} --
Minimising Eq.~(\ref{eq:chisq}) using initial parameters that are close to the ``true'' values is crucial 
to avoid instabilities. We produce our initial parameters using a combination of effective mass computations 
and fits to a different number of states. The mass of the $n^{\rm th}$ state can be estimated using the 
effective mass if knowledge about the parameters of the lighter $n-1$ states is available. To do so, we 
subtract the model composed of the $n-1$ states from the correlation function estimate, eliminating those 
contributions from the correlation function; see Eq.~(\ref{eq:specdec}). We can use this procedure
iteratively to estimate the masses of the wanted excited states. Afterwards, one should always fit a
one-state model to the subtracted correlation function to produce a more robust estimate of the initial
parameter. As errors propagate, this technique is less reliable for higher order excited states.

\begin{figure}[t]
 \centering
    \includegraphics[scale=0.2]{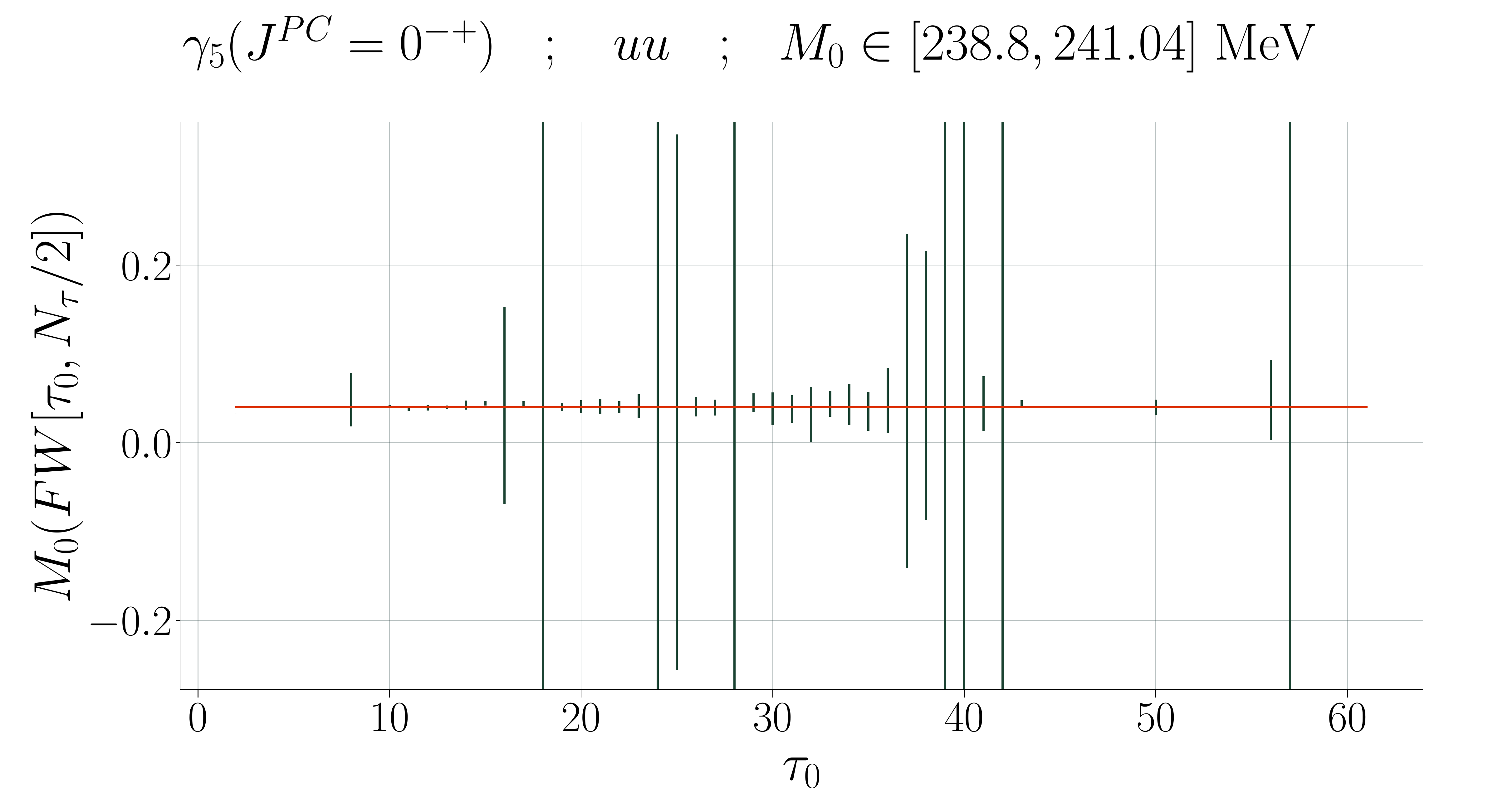}
    \caption{
        Ground state mass of the \textit{local-local} pion correlation function ($\gamma_5, uu$) as a 
        function of the starting time $\tau_0$ defining the fit
        window used in the regression; the
        fit windows used contain all points in the region $[\tau_0, N_\tau/2]$. The error bars correspond
        to the estimate of the ground state mass at a given time window. In some cases the error bars are too small to be visible. Error bars change rapidly in regions where additional excited states become relevant.
        The line corresponds to the median
        mass using all fit windows; this result is independent of the fit window used. The temperature
        of the system is $T = 46$~MeV ($N_\tau = 128$).
    }
   
\end{figure}

\subsection{Regression at fixed fit window and extraction of the final ground state mass}

We will now explain how regression is carried out using a fixed fit window. As we do not know when
excited states become relevant, we should include models with a different number of states. For each
of the models, we can produce an estimate of the ground state mass. At the end, we need to select the best
estimate of the ground state mass at the current fit window. To obtain this value, we use the so-called
\textit{corrected Akaike Information Criterion} (\textit{AICc}) \cite{Akaike,Akaike:1998zah}. The 
$AICc$ can be used to measure the relative likelihood of the data description between two models; 
the model with the lowest $AICc$ is the most likely to describe the data. It can be defined using

\begin{equation}
    AICc = N_\theta - \log(\hat{L}) + \frac{N_\theta^2 + N_\theta}{N_{\rm fw} - N_\theta - 1},
\end{equation}
where $N_\theta$ is the number of parameters in the model, $\hat{L}$ is the likelihood function
evaluated at the estimated parameters and $N_{\rm fw}$ is the number of points included in the fit.

Using the $AICc$, we can compute the relative likelihood between two models $m_1$ and
$m_2$,
\begin{equation}
    l(m_1, m_2) = \exp\left( -\frac{1}{2} [ AICc(m_1) - AICc(m_2)]\right).
\end{equation}
This quantity measures how likely $m_2$ is to describe the data compared to $m_1$. Provided we set 
$m_1$ to the model with the lowest $AICc$ among all models available, then we can use the values 
of $l$ to measure the \textit{relative model quality}. We can utilise this information to compute 
the best estimate of the mass at the given time window by calculating the weighted average of all 
masses; we use $l$ as weights. This technique allows us to promote the influence of high-quality 
models in our result while avoiding manually discarding any models. Finally, the error in the weighted 
mass can be extracted using a bootstrap analysis.

For each fit window used in the fit, we collect an estimate of the ground mass $M_0(FW)$. Our final
estimate of the mass, independent of the fit window, is extracted using the median of all $M_0(FW)$.
We use the median as it is a robust statistic; outliers tend to be present due to the unstable
nature of correlated fits. Confidence intervals on the median estimation can be constructed using
bootstrap; confidence intervals are not guaranteed to be symmetric.

%% file: sections/results.tex
\section{Results}

\begin{figure}[t]

    \centering
    \begin{minipage}{0.5\textwidth}
        \centering
        \includegraphics[width=1.0\textwidth]{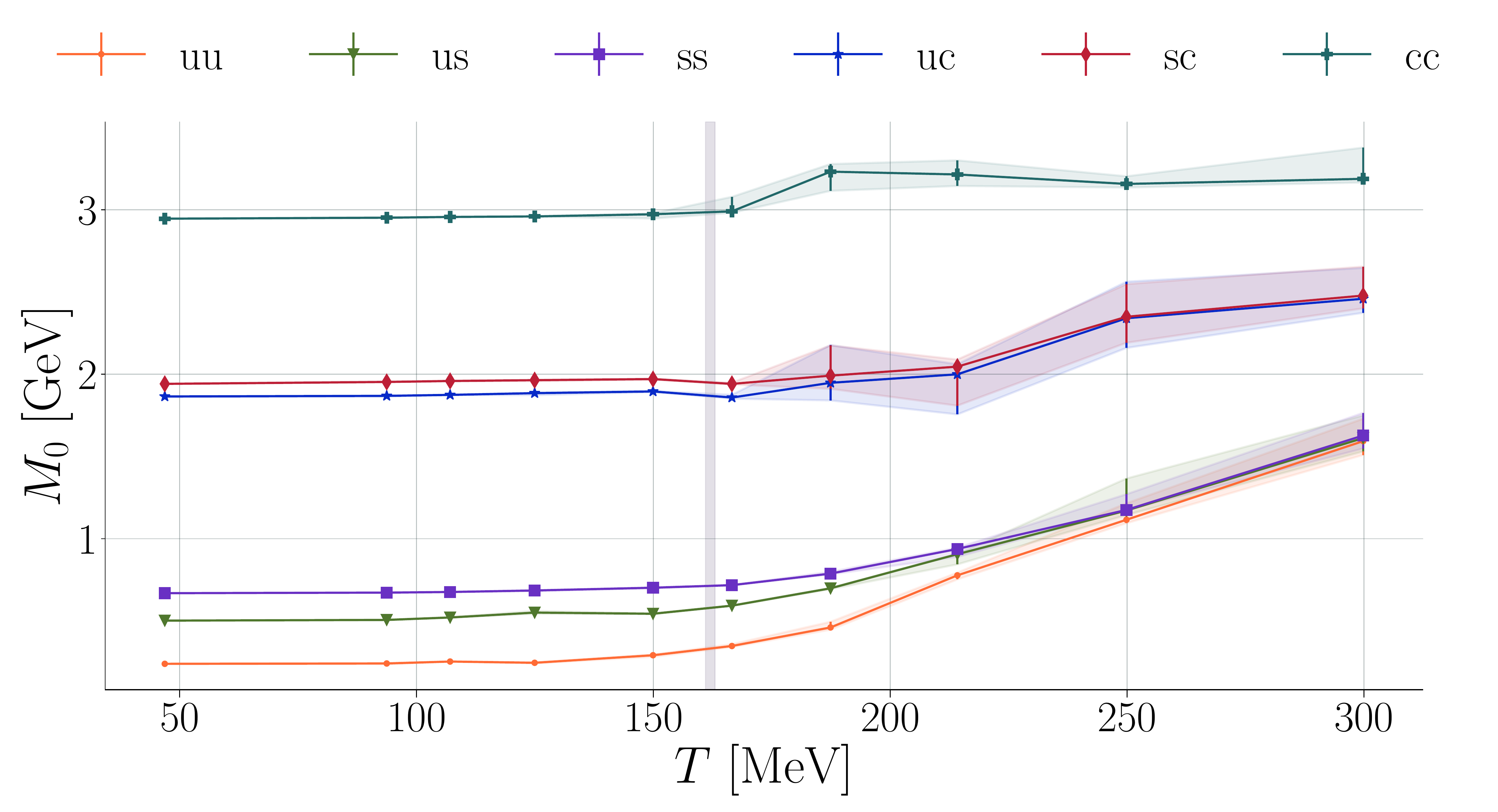}
        \caption*{Pseudoscalar channel}
    \end{minipage}\hfill
    \begin{minipage}{0.5\textwidth}
        \centering
        \includegraphics[width=1.0\textwidth]{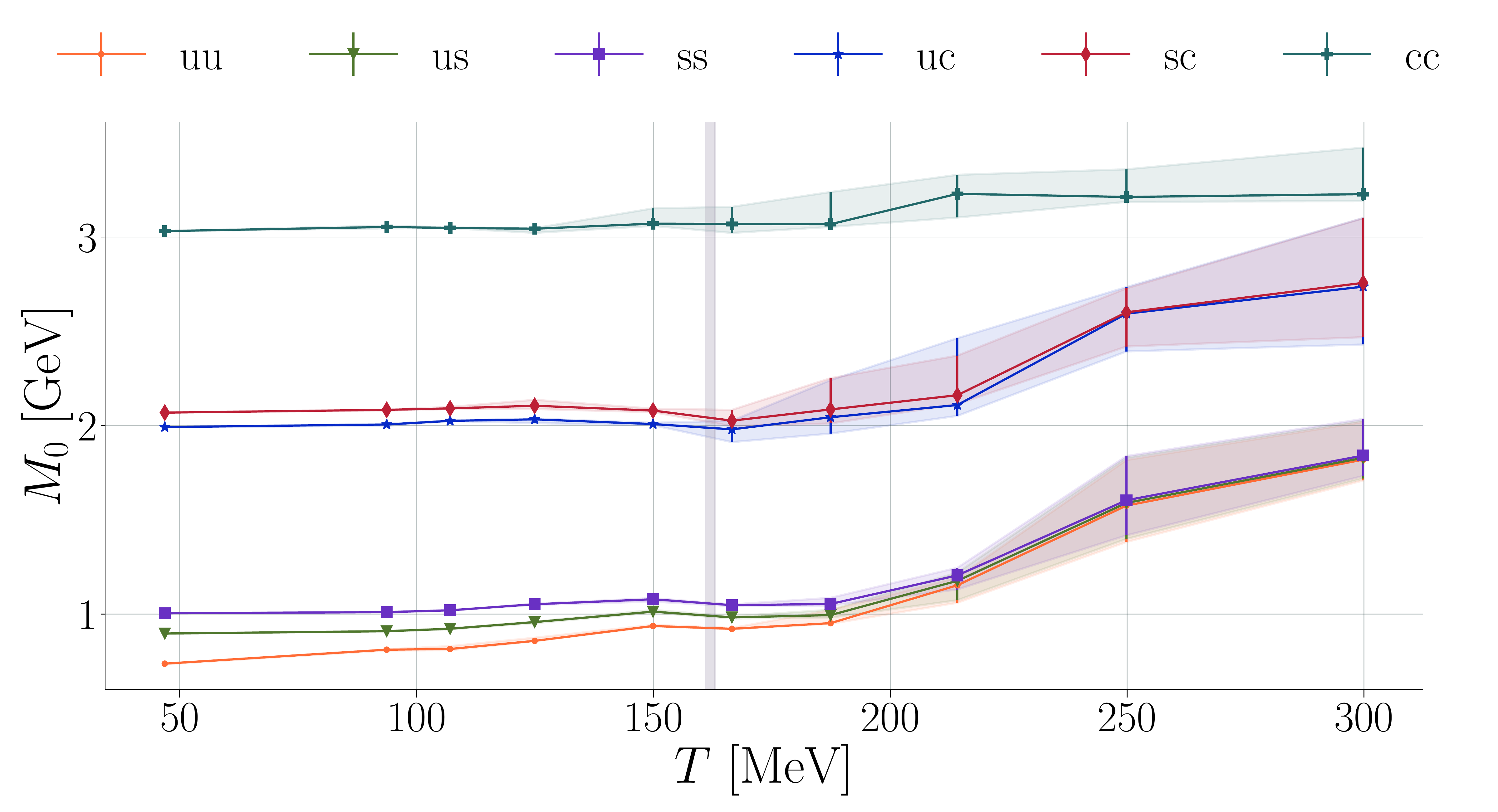}
        \caption*{Vector channel}
    \end{minipage}
    \caption{
        Temperature dependence of the ground state masses in the pseudoscalar ($\gamma_5$) and the
        vector ($\gamma_\mu$) channels. The vertical gray line corresponds to the pseudocritical temperature
        $T_c$.
    }
        \label{fig:g5gidep}
\end{figure}

We will now present some results for the ground state masses for different states. Fig.\ \ref{fig:g5gidep}
contains the temperature dependence of the different flavour combinations in the pseudoscalar and vector
channels. Two different trends can be seen. In the low-temperature regime, where the system is in the
\textit{hadronic phase}, the masses show minimal temperature dependence. For the light quarks, some 
dependence can be seen in the vector channel. On the other hand, above the pseudocritical temperature 
($T_c\sim 164$ MeV), which is estimated using the inflection point of the chiral condensate, where the 
system is a quark-gluon plasma, the masses vary with the temperature, when the results are taken at face
value. This effect is stronger in the light sector ($uu$, $us$ and $ss$). For the heavier sector, involving
charm, the effect is milder and the charmonium $cc$ states are almost unaffected by thermal effects. In the
high-temperature regime, the masses of the light mesons increase until becoming degenerate. The mass increase
and the degeneration of the light flavour combinations can be explained by thermal effects: the inherent
light-quark-energy scales are smaller than the temperature scale and the system is dominated by thermal
excitations. Note that the uncertainties in the estimates grow with temperature due to the combined fact 
that the model used in the fit is not expected to be accurate as the temperature increases (to be discussed
next) and the fact that the number of points available in the fit decreases with $T$.

The apparent presence of thermal effects for light mesons in the high-temperature regime leads to a 
natural question: is the spectral decomposition underpinning Eq.~(\ref{eq:specdec}) valid at high $T$? 
The answer is clearly no: at high temperature, thermal effects, collective excitations and screening 
are expected to impact mesonic correlators. Light meson states will disappear (deconfinement) and although
states involving charm may survive in the quark-gluon plasma, they are expected to be no longer describable 
by delta functions/simple poles. The inability of Eq.~(\ref{eq:specdec}) to represent the high-temperature
regime generates a problem: the quantities extracted in this region cannot be interpreted as masses. This
problem is more severe in the light sector, where thermal effects are more dominant. Nevertheless, the 
results contain meaningful information about the status of the system in the sense that degeneracies and
symmetry restoration can be studied from the outcome of the regression analysis, as it gives information on
the underlying correlators, irrespective of the interpretation as ground state mass. 

\begin{figure}[t]
    \centering
    \includegraphics[scale=0.2]{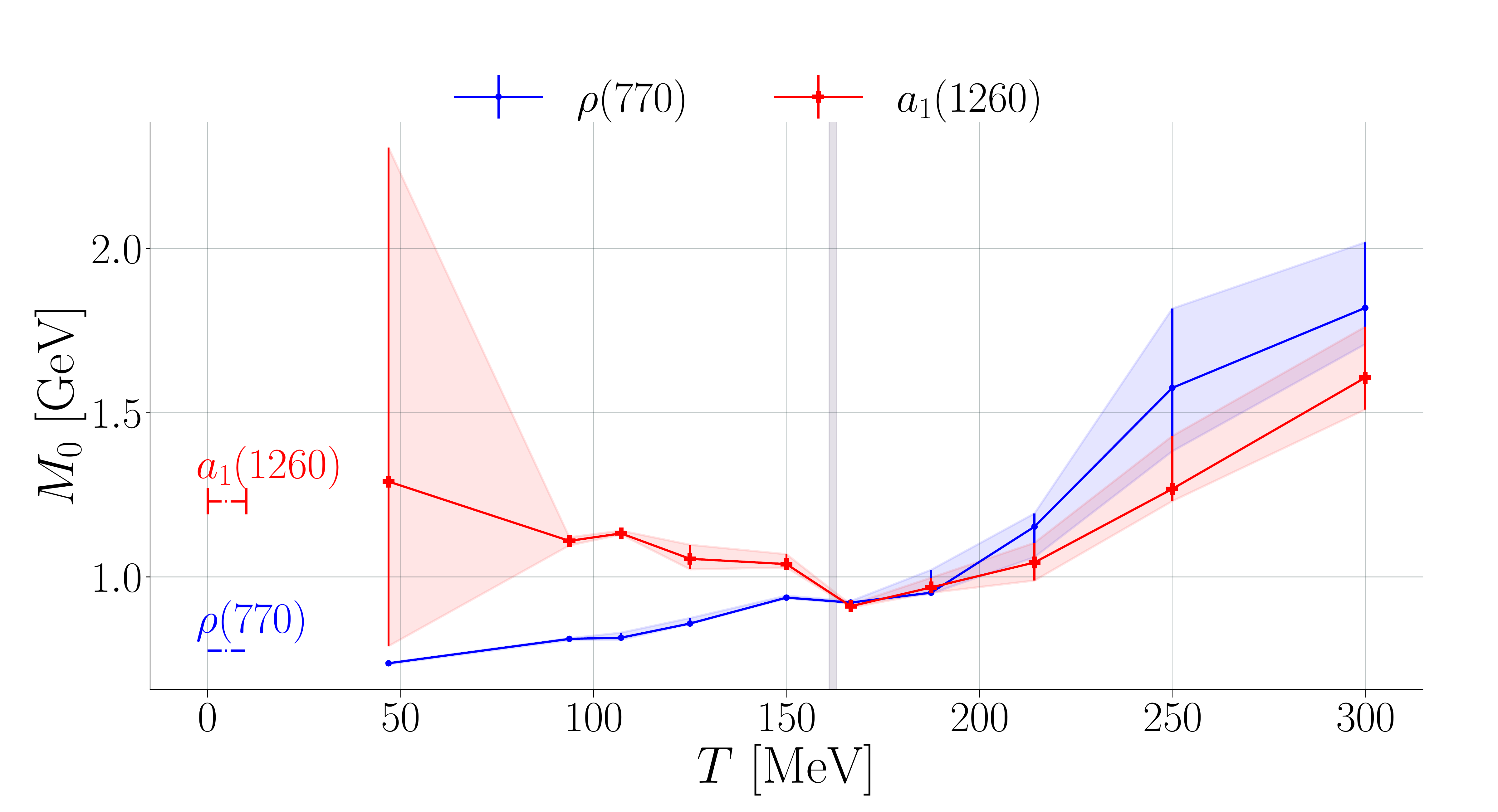}
    \caption{
        Temperature dependence of two $SU(2)_A$ related channels: vector -- $\rho(770)$ --  and axial-vector -- $a_1(1260)$.  
        The estimate of the correlation function in the axial-vector channel at the lowest temperature is remarkably noisy, affecting the confidence intervals.
    }
    \label{fig:SU2A}
\end{figure}

We demonstrate this in more detail in the context of SU(2)$_A$ chiral symmetry restoration and the 
temperature dependence of states related by this symmetry. Here we discuss the vector and axial-vector
channels in the lightest flavour combination, connected to physical states $\rho(770)$ and $a_1(1260)$
respectively, see Fig.\ \ref{fig:SU2A}. At low temperature, the states are non-degenerate. However, at higher
temperature, they are expected to become degenerate as the SU(2)$_A$ symmetry is restored. The results of our
simulations in Fig.\ \ref{fig:SU2A}  appear to confirm this expectation, with the degeneracy emerging at or
near the transition temperature.

\begin{figure}[t]
        \centering
    \begin{minipage}{0.5\textwidth}
        \centering
        \includegraphics[width=1.0\textwidth]{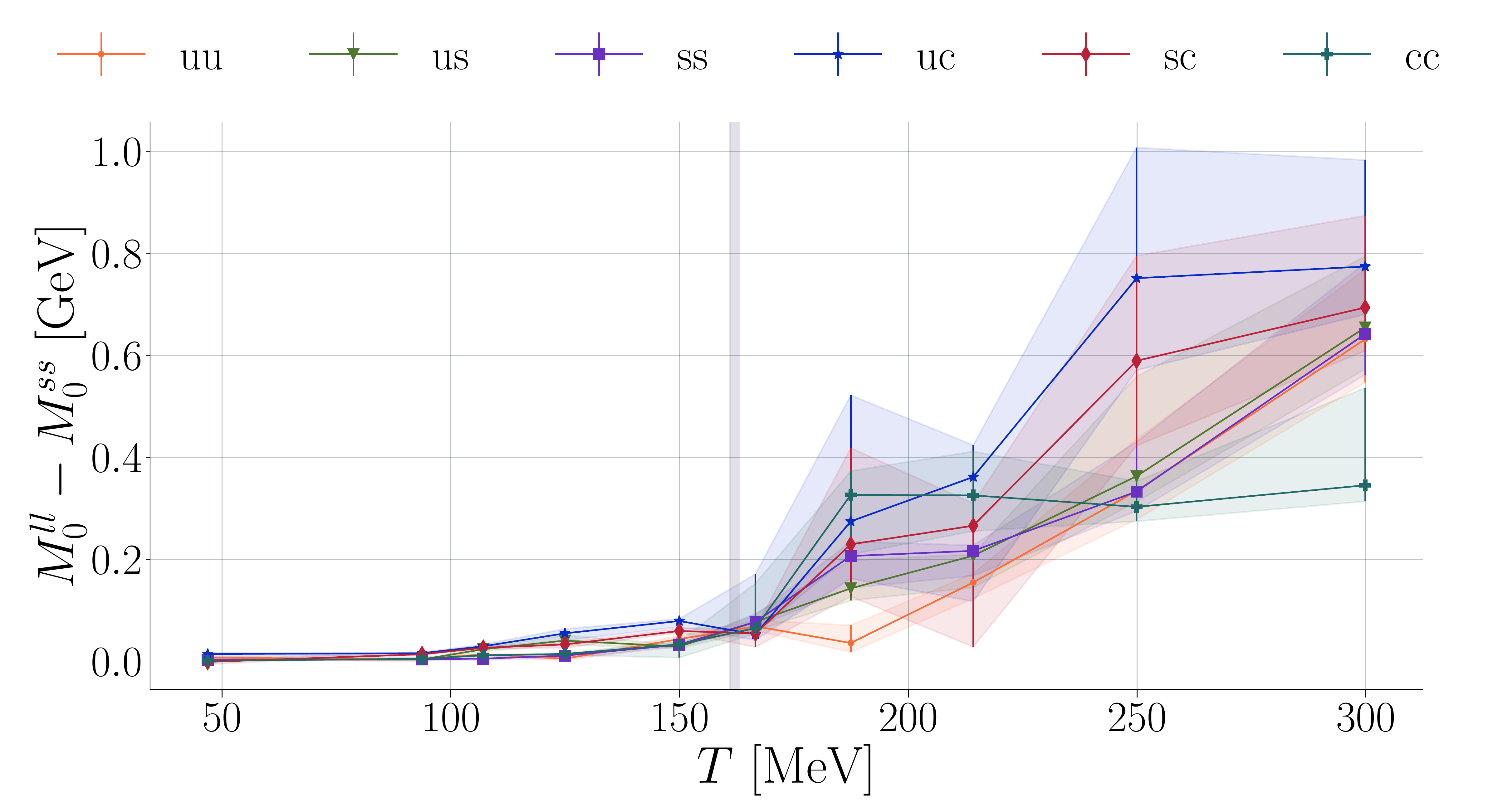}
        \caption*{Pseudoscalar channel}
    \end{minipage}\hfill
    \begin{minipage}{0.5\textwidth}
        \centering
        \includegraphics[width=1.0\textwidth]{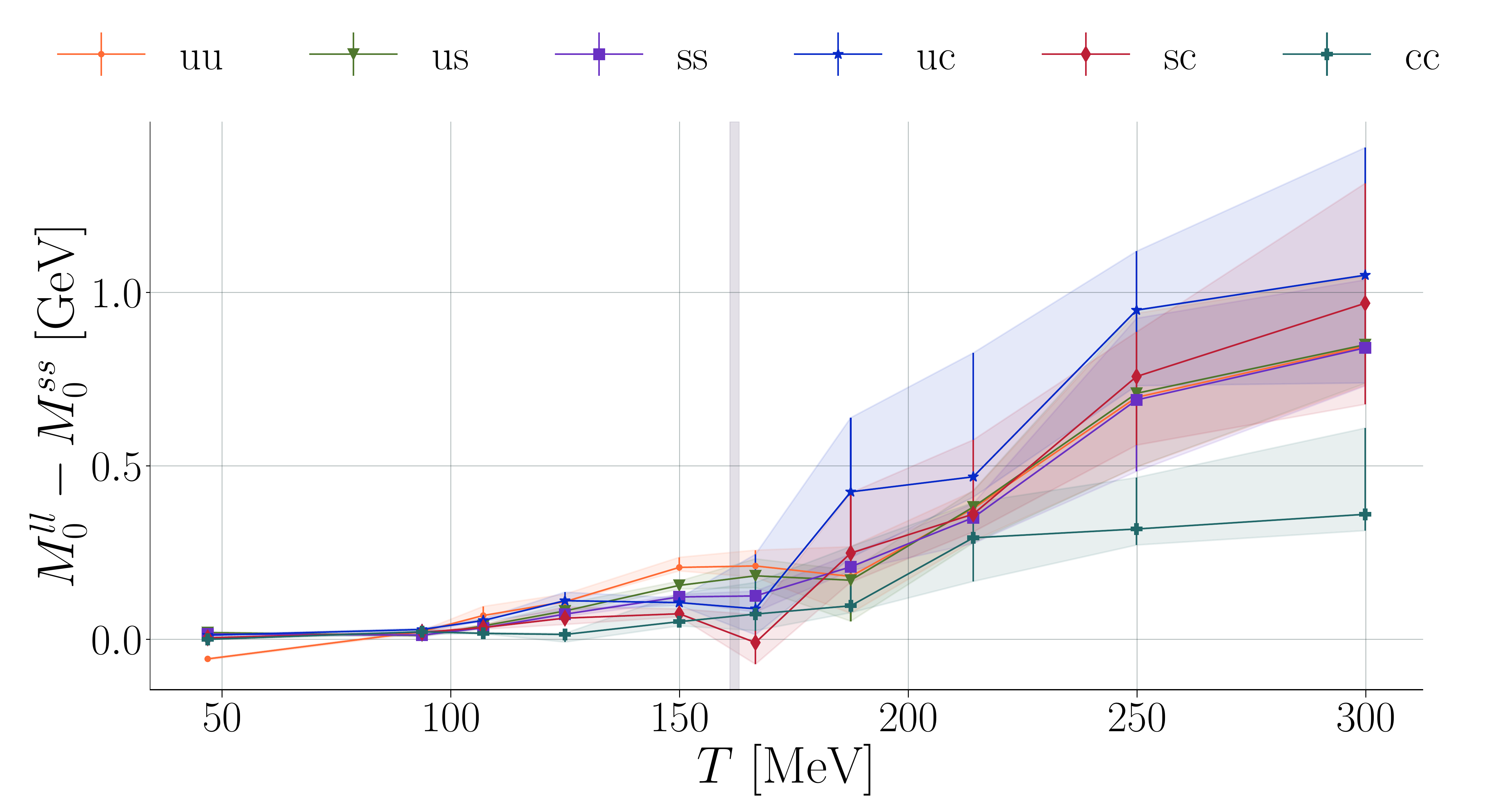}
        \caption*{Vector channel}
    \end{minipage}
    \caption{
        Temperature dependence of the difference between the ground state masses extracted using local and
        smeared sources, $M_0^{l-l}-M_0^{s-s}$, in the pseudoscalar ($\gamma_5$) and vector ($\gamma_\mu$)
        channels for all flavour combinations.
    }
        \label{fig:llss}
\end{figure}

Finally, in Fig.\ \ref{fig:llss} we present some results comparing the masses extracted using different 
types of  sources: local and smeared ones. Smearing is designed to decouple the excited states in a
correlation function. As a consequence, the ground state signal is enhanced, thus leading to a better
estimate of the ground state mass. Where the ground state is well-defined, local and smeared ground state mass estimates should be equal. In the hadronic phase, our results indeed are comparable. However, in the deconfined phase the
estimates are no longer equal. The implication is that the single-pole Ansatz given in Eq.\ (\ref{eq:specdec}) is not valid and the ground state is no longer present.
The larger errors at high temperature are another indication the fits are no longer working.

%% file: sections/conclusions.tex
\section{Conclusions}

The regression analysis presented allows the estimation of ground state masses in a systematic manner.
However, as stated before, these masses are not reliable at all temperatures. In the hadronic phase, the
Ansatz consisting of a sum of simple poles, see Eq.~(\ref{eq:specdec}), is sufficient to obtain reliable
estimates. A minimal temperature dependence of the masses in this regime is observed. In contrast, as the
temperature increases, thermal effects dominate and Eq.~(\ref{eq:specdec}) is no longer valid. More complex
models and algorithms would be needed to fully capture the information present in the correlation functions 
at high temperature, reconstructing e.g.\ the spectral functions directly. Although the approach used here 
is not completely accurate, pushing the boundaries of this established method is not useless. Information
about degeneracies and symmetry restoration at the level of the correlation functions can be extracted using
simple regression, and those results can be used as a firm ground on which more complex analysis tools are
built.

A remarkable finding of our analysis is the degeneracy of SU(2)$_A$ related states. Our results confirm 
the non-degeneracy of the $\rho(770)$ and the $a_1(1260)$ states in the hadronic phase. As the temperature
increases, the states get closer and become degenerate at or around the transition temperature to the
quark-gluon plasma and remain degenerate at higher temperatures. 

As a final important exercise, we compared correlators obtained using local and smeared sources. An agreement
between the extracted ground state masses provides an important boost for the validity of the approach. This
is observed in the hadronic phase. On the other hand, likely reasons for a discrepancy at high temperature are that the ground
state is no longer given by a simple pole or no longer clearly discernible, due to thermal effects and/or
deconfinement. This was seen in the quark-gluon plasma, in all flavour combinations and channels studied. 
This leaves the question of the impact of smearing at high temperature; further research on the effects of
smearing on high-temperature observables is therefore needed.

\newpage
\noi
 {\bf Acknowledgements} -- 
This work is supported by the UKRI Science and Technology Facilities Council (STFC) Consolidated Grant 
No.\ ST/P00055X/1 and ST/T000813/1. We are grateful to DiRAC, PRACE and Supercomputing Wales 
for the use of their computing resources and to the Swansea Academy for Advanced Computing for support. 
This work was performed using the PRACE Marconi-KNL resources hosted by CINECA, Italy and the DiRAC 
Extreme Scaling service and Blue Gene Q Shared Petaflop system at the University of Edinburgh operated 
by the Edinburgh Parallel Computing Centre. The DiRAC equipment is part of the UK's National e-Infrastructure 
and was funded by UK's BIS National e-infrastructure capital grant ST/K000411/1, STFC capital grants
ST/H008845/1 and ST/R00238X/1, and STFC DiRAC Operations grants ST/K005804/1, ST/K005790/1 and ST/R001006/1.